# Light Control in a Hemicylindrical Whispering Gallery Microcavity-Parallel Plate Waveguide System

Henrik Parsamyan, Hovhannes Haroyan and Khachatur Nerkararyan*

Department of Microwave Physics and Telecommunication, Yerevan State University, Yerevan, Armenia

*Corresponding author.

Address: Republic of Armenia, Yerevan, 0025, 1 Alex Manoogian
E-mail: knerkar@ysu.am

We have shown that a structure composed of a semiconductor hemicylinder and a dielectric waveguide separated by a thin metal layer permits efficient and easy light control. In such a system, it is possible to ensure matching between TE and TM modes of a low refractive index waveguide with the corresponding whispering gallery modes of the hemicylinder of a higher refractive index, a circumstance that allows one to reduce the resonator sizes to the order of the exciting wavelength in the vacuum still having rather high $Q$-factor $\sim 2\cdot 10^4$. Due to pronounced electro-optical properties of the resonator medium, one is able to effectively control the output light intensity by varying the refractive index of the hemicylinder in the order of $10^{-4}$.

## 1. Introduction

Optical whispering gallery mode (WGM) microresonators, representing geometries with spherical and cylindrical symmetry, have been widely investigated as powerful tools for a wide range of fields, including cavity quantum electrodynamics[1], sensors[2], lasers[3] and photonic filters[4] owing to the combination of quality $Q$-factors as high as tens and hundreds of millions and low modal volumes[5–7]. One of the aspects to achieve such high values of the $Q$-factors is large sizes of resonators, usually exceeding the excitation wavelength in the vacuum tens and hundred times. One way to reduce the resonator sizes, while preserving its high-$Q$ characteristics, is using high refractive index (RI) materials as a resonator host medium. Analysis of the excitation of WGM resonators by the evanescent coupling shows that effective excitation can be achieved when the RI of the resonator medium matches to that of the exciting fiber core or prism[8–10], usually made of conventional glasses with refractive index $\sim 1.4$-$1.6$[11]. This means that for high RI WGM resonators (e.g. high index glass, crystalline or III-V semiconductor resonators), enabling to reduce the resonator sizes while keeping high $Q$-factors and exploiting the advantages of strong electro-optical properties of such materials[7], one can either use fiber tapers with the high RI core (close to the RI of the resonator) or develop new approaches to ensure efficient excitation[12]. The fabrication of high RI core fibers, on the one hand, is challenging and, on the other hand, their use is limited due to the fragility of tapered regions, as well as sensitivity of the coupling efficiency from the gap between the tapered fiber and the cavity, needing for devices to precisely adjust the nanometric gap width of the coupling region.

In fact, from a practical standpoint, on-chip cavities with integrated input/output bus waveguides are used, again utilizing the concept of the evanescent coupling, and having an important role in the fast-growing integrated photonics. Depending on the resonator-waveguide positions, the lateral (the waveguide and the cavity are on the same planes) and vertical (the waveguide and the cavity are on different planes) coupling methods are distinguished[13,14]. The nanometer-scale gap of the coupling region in such systems is etched by using ultrahigh-resolution lithography.

The plane-wave excitation of WGMs in a high RI hemicylindrical microresonator based on a dielectric-metal-dielectric structure was investigated in the manuscript [15]. The system consisted of a dielectric substrate covered by a thin metal layer, onto which was placed a GaAs hemicylinder. The analysis suggested that the $Q$-factor of the GaAs resonator with $R = 3$ µm reaches up to $2.5 \cdot 10^4$. In addition, the energy stored in the half-cylinder strongly depends on the change of its RI about the order of $10^{-4}$. Although the excitation method of the configuration described above is very easy and the $Q$-factor is reasonably high, the main drawback of the configuration is related to the detection of the resonance positions requiring to use of additional devices such as a sharp-tipped filament of fiber[16] or to measure the back-radiated energy from the metal layer. This issue can be solved by modifying the structure and using the concept of the common waveguide-based excitation allowing one to detect the resonances via measuring the transmission directly from the output of the waveguide, at the same time keeping the advantages of the excitation technique of the studied configuration.

In this paper, we suggest a system consisting of a hemicylindrical microcavity - dielectric waveguide coupled via a thin metal layer. The resonant positions of such system can be easily defined by measuring the transmitted power on the output of the waveguide. Using of a high RI material as the resonator host medium allows one to reduce the resonator sizes to the order of the excitation wavelength in the vacuum. At the same time, high RI semiconductors are materials characterized by large electro-optical and nonlinear coefficients having an essential role in efficient light control.

## 2. Results and discussion

The structure is composed of a waveguide with dielectric core sandwiched by metal layers from the top and bottom, moreover, the thickness of the top metal layer is few tens of nanometers. The resonator of the hemicylindrical form and with relatively high RI is placed on the top of the waveguide (see Fig. 1). $h$ is the metal thickness, $d$-waveguide width, $R$-the radius of the hemicylinder and $n_s$-its index of refraction. The red arrow on the left side of the waveguide indicates the input power and the blue arrow-output power.

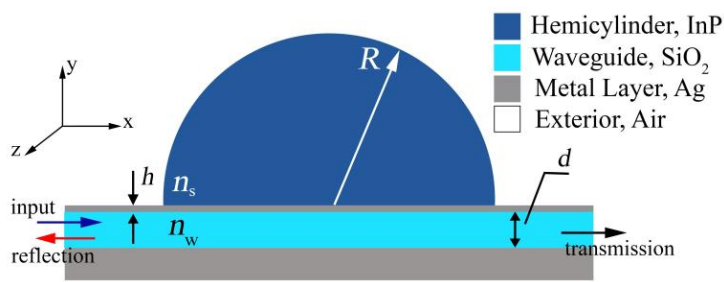

Fig. 1. The schematic of the structure. $h$ is the thickness of the metal between the hemicylinder and waveguide, $d$ is the width of the waveguide and $R$-the hemicylinder radius. The exterior medium of the hemicylinder is air.

To investigate the resonator, we performed a full-wave numerical analysis based on the finite element method using the following parameters for the system geometry: $d = 500$ nm, $h = 50$ nm and $R = 3$ µm for waveguide width, the thickness of a metal layer and hemicylinder radius, respectively. Throughout simulations, as the metal we used Ag[17], and for the resonator and waveguide materials, InP[18] and $SiO_2$[19] were used, respectively. The system is surrounded by air ($n_e = 1$).

It is noteworthy that due to the small thickness of the metal layer waveguide modes propagate by multiple total reflections ($n_w > n_e$), thereby the presence of such a thin metal film does not lead to the leakage of the wave energy and the wave propagation is accompanied by only relatively small Joule losses. However, the situation is different, when the wave passes through the metal-resonator region. Since the condition of the total reflection is not satisfied (in this case $n_w < n_s$), the wave penetrates into the hemicylinder, where at a resonant wavelength WGMs are formed, whereas at nonresonant wavelengths about 20 % of the wave energy is lost while the wave propagates by the waveguide (see Fig. 2). Also, worth noting that in this case, the coupling is similar to that of an incident plane wave on a metal film to surface plasmon polaritons (SPP) in the Kretschmann configuration[20]. Numerous experiments related to the excitation of SPPs show that the thickness of 50 nm of a metal layer is optimal for a coupling.

Normalized transmission (black) and reflection (red) spectra of the waveguide-hemicylinder system for fundamental transverse electric (TE) and transverse magnetic (TM) waveguide modes in the spectral range of 1000-1050 nm are shown in upper and lower plots of Fig. 2(a), respectively. A series of dips in the transmission spectrum, as well as peaks in the scattering spectrum, caused by a waveguide mode coupling to the microresonator, correspond to the resonances of the hemicylinder. Labels ($\ell, m$) of both TE and TM modes stand for the radial $\ell$ and azimuthal $m$ mode numbers. The electric field distribution inside the hemicylinder can be defined by the following equation [15]

$$E_z(r,\varphi,t) = A J_m(k_s r) \cdot \sin(m\varphi) \cdot \exp(i\omega t) \qquad (1)$$

where A is a constant, $J_m(k_s r)$ is the Bessel function of the first kind and $m$-th order (azimuthal quantum number), $k_s = 2\pi n_s / \lambda_0$ is the wavenumber, $n_s$ is the hemicylinder RI and $\lambda_0$ is the free-space wavelength of the exciting wave.

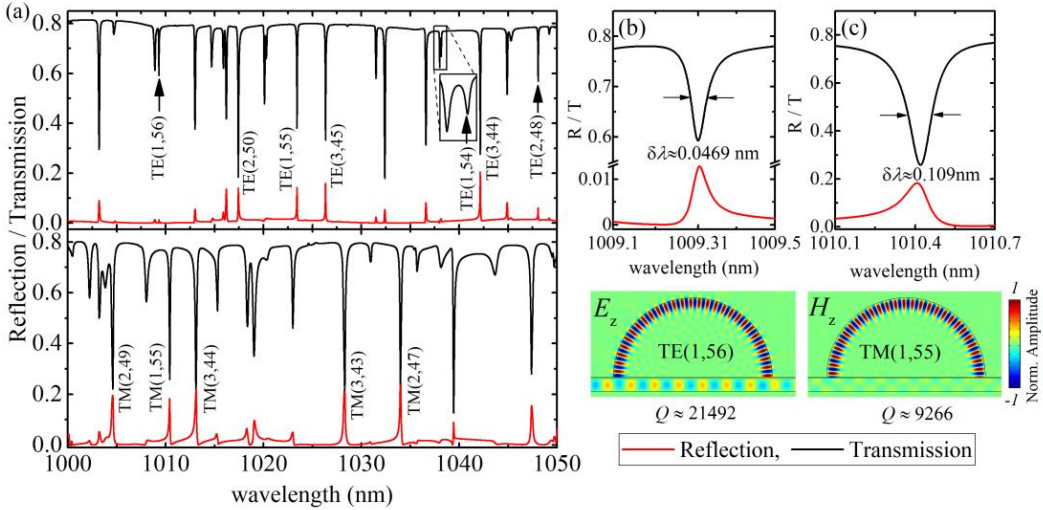

Fig. 2. (a) Transmission (black) and reflection (red) spectra of the InP hemicylindrical microresonator excited by a TE (upper) and TM (lower) waveguide modes. The same for a (b) TE(1,56) and (c) TM(1,55) modes with the corresponding field distributions below. The first ($\ell$) and second ($m$) numbers in mode notation correspond to the radial and azimuthal quantum numbers of a WGM. Arrows in (b) and (c) show the FWHM $\delta\lambda$ of the resonant transmission curves. Note that the vertical axis of the (b) is broken to scale the reflection spectrum.

The radial quantum number $\ell$ indicates the quantity of extrema of the Bessel function $J_m(k_s r)$ on the length $R$ along the radial $r$ direction. Note that the expression similar to (1) can be written for magnetic field distribution inside the resonator.

One considers the hemicylinder and the metal layer on its polished surface as a Fabry-Perot-like (FP) cavity, where the metal layer represents both mirrors and the wave propagates by a curved boundary. At the resonant, a part of energy in a waveguide penetrates to the resonator through a thin metal layer. The WGMs then are formed by multiple total internal reflections from the resonator-air boundary and reflections from the metal layer. Once a resonant WGM is established, the resonator losses are mainly conditioned by Ohmic losses in the metal (Ag) layer and the back transmitted energy from the resonator to the waveguide. It is important to note that at the resonant regime losses from the hemicylinder curved boundary are quite small and can be neglected.

In Fig. 2(a) we indicated three types of TE and TM WGMs with $\ell = 1, 2$ and 3 radial orders. Hereinafter, a fundamental WGM is called the one with radial number $\ell = 1$, corresponding to the modes with smallest mode volumes. Transmission and reflection spectra of particular TE(1,56) and TM(1,55) fundamental modes with respective field distributions are presented in Fig. 2 (b) and (c), respectively. The full width of the half-maximum (FWHM) of TE(1,56) mode $\delta\lambda \approx 0.0467$ nm, while that of TM(1,54) mode is about 2 times larger and equals $\delta\lambda \approx 0.109$ nm. $Q$-factors of the resonator are evaluated by the general expression $Q \approx \lambda_0/\delta\lambda$, where $\lambda_0$ is the resonant wavelength and $\delta\lambda$ is the FWHM of the transmission curve. Thus, the calculated total $Q$-factors of TE(1,56) and TM(1,54) fundamental modes are $\sim 2.14 \cdot 10^4$ and $\sim 9.27 \cdot 10^3$, respectively. Free spectral range of the resonator, representing the spectral width between two modes with $m$ and $m+1$ azimuthal and the same radial $\ell$ numbers - $\Delta\lambda_{FSR} = \lambda_{\ell,m} - \lambda_{\ell,m+1}$, is about $\sim 14$ nm for TE modes and $\sim 15$ nm for TM modes. Analyzing Fig. 2(a) one finds out that the resonant processes are accompanied by significant losses. Practically for all modes the sum of the transmission and reflection noticeably smaller than the unit. Since the $Q$-factor of the resonator is relatively high, one supposes that energy losses take place outside of the hemicylinder. As already noted, due to violation of the total reflection condition in the region "under hemicylinder" of the waveguide channel energy leakage occurs. At the nonresonant regime, overall leakage is about $\sim 20\%$ (see Fig. 2a). A similar leak also occurs in the case of a resonant propagation regime, since both the incoming wave and the waves emanating from the resonator partially pass through the indicated leakage region. Analysis reveals that losses in the hemicylinder from the metal layer (energy leakage to the waveguide) can be reduced by increasing the metal thickness. However, this will yield to decrease of the depth of resonant dips in the transmission spectrum.

In addition to above mentioned modes with the smallest mode volume, other resonant TE modes exist with relatively high-$Q$-factors and larger mode volumes (i.e. larger radial numbers) (see Fig. 3(a) and 3(b)). For instance, the $Q$-factors of the TE(2,48) and TE(3,45) modes are about $\sim 22087$ and $\sim 17234$, respectively, whereas for the TM(2,46) mode $Q \sim 4628$.

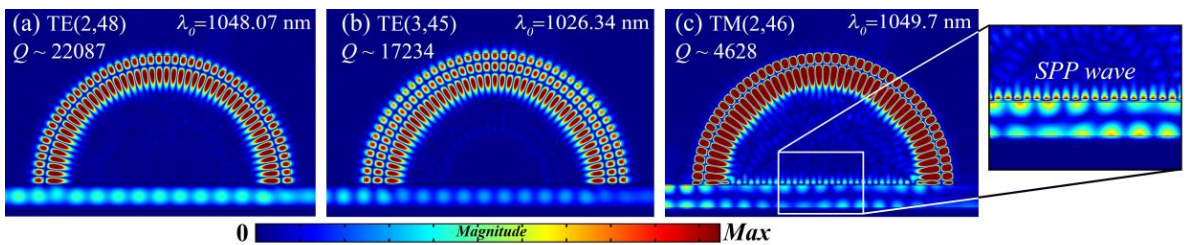

Fig. 3. (a)-(b) Electric field distribution of the TE(2,48) and TE(3,45) resonant modes. (c) Magnetic field distribution of the TM(2,46) mode. The zoomed frame shows *SPP*s at the metal-hemicylinder boundary. All modes are presented with corresponding $Q$-factors and resonant wavelengths. The resonator medium is InP.

Calculations reveal that the $Q$-factors of the TM modes are rather lower than that of TE modes. The reason for such difference can be understood from Fig. 3(c), showing the presence of the SPPs on the metal-hemicylinder boundary, which reasonably increases the energy losses and thus reduces the $Q$-factors of TM modes.

From the standpoint of the technical implementation, a resonator shall retain its common characteristics (for instance relatively high $Q$-factor) even for lower structural symmetry. This problem has been studied by modifying the symmetry of the transverse cross-section by using a hemiellipse instead of a hemicylindrical resonator. Figure 4(a) depicts the transmission (black) and reflection (red) spectra of hemielliptical resonator with semi-minor axis $b = 2.5$ µm and semi-major axis $a = R = 3$ µm, schematically displayed in the top of Fig. 4(b).

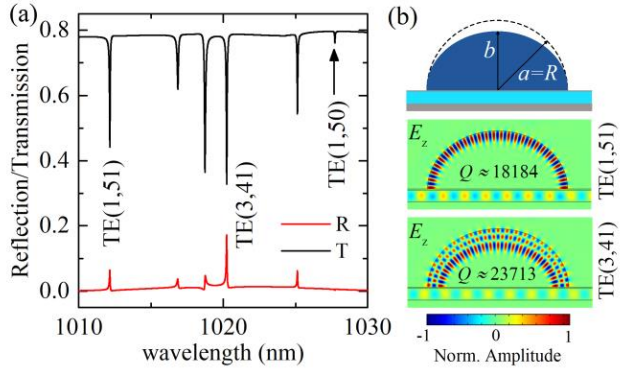

Fig. 4. (a) Transmission (black) and reflection (red) spectra of the hemielliptical microresonator. (b) The schematic of the configuration and $E_z$ field distribution of the TE(1,51) and TE(3,41) modes with corresponding Q-factors. The major axis $a$ of the is equal to $R=3$ μm, while the minor one $b = 2500$ nm. InP is used as a resonator medium.

In such a structure the $Q$-factor of the resonator is comparative with that of the hemicylindrical, however, the number of resonant modes, as well as the azimuthal order of modes decrease. The preceding features are due to a partial violation of the symmetry of the structure. Besides in the present case, the effective optical path of the cavity decreases ($b < a = R$). Two resonant modes of the hemielliptical system are shown in the bottom of Fig. 4(b) with the following characteristics: TE(1,51) at 1012.16 nm with $Q \sim 18184$ and TE(3, 41) at 1020.24 nm with $Q \sim 23713$.

To show the prospects of using such a structure as a modulator, in Fig. 5 we display the dependence of the transmission at the output of the waveguide on the resonator RI change for modes TE(1,56), TE(2,50) and TM(1,55) at resonant wavelengths 1009.31nm, 1017.43 nm and 1010.42 nm, respectively. Note that RI of the InP is around 3.31 near the wavelength 1010 nm.

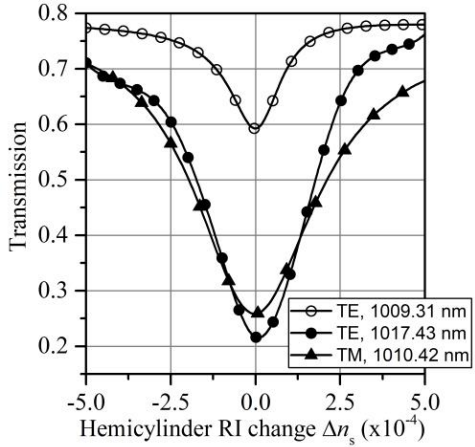

Fig. 5. Dependence of the transmitted intensity at the output of the waveguide on the refractive index change of the hemicylinder for resonant modes TE(1,56) at 1009.31nm, TE(2,50) at 1017.43 nm and TM(1,55) at 1010.42 nm.

Although there are certain analogies between the studied configuration and the FP resonator, in terms of the transmission of the wave the results are totally different. Here, the deviating from the resonant wavelength results in a sharp increase in the transmission (see Fig. 2 and Fig. 4). As a consequence, the transmission increases also upon shifting from the resonance through varying the RI of the hemicylinder (Fig. 5). The fact is that similarly to the FP resonator, here the hemicylinder is not excited at non-resonant wavelengths and the metal layer has a role of a fairly good mirror such that the wave passes through the structure with relatively small losses. However, at resonant along with the waveguide channel, the "resonator channel" (i.e. the inner curved boundary of InP-

Air) also appears. In this case, the interference between the waves of two connected channels drastically decreases the transmission.

Hence, the intensity of the transmitted light can be controlled by varying the RI of the resonator by the order of $10^{-4}$, which can be easily achieved by the large variety of the materials with noticeable electro-optical properties, such as III-VI semiconductors.

The issue of fabrication of the structure under study is also important. The fabrication process of such a structure, in our view, can be divided into two parts. First, to create a waveguide structure, metal and dielectric layers should be applied to the semiconductor layer. Then, to strengthen the structure, it is necessary to apply a sufficiently thick layer of the substrate. At the second stage, only after the thickness of the semiconductor layer is reduced to the required size, a hemicylinder can be created. The fabrication process is similar to the well-introduced technique of the fabrication of semiconductor lenses [21–23].

## 3. Conclusion

Thus, mode-matching of the whispering gallery modes in the hemicylindrical high refractive index microcavity and low refractive index dielectric waveguide TE and TM modes, realized by a thin metal layer between the components, ensures efficient control of the transmitted output light. Such a process is based on interference of the light propagating by two connected paths, into one of which the light energy is increased resonantly. Using a high refractive index material−InP as the resonator host medium one is able to reduce the resonator sizes to the order of the excitation wavelength in the vacuum, however with still high $Q$-factor ~ $2.1 \cdot 10^4$. Such semiconductors are featured by large values of electro-optical and nonlinear coefficients thus supporting efficient and easy control of the transmitted light even by varying the refractive index of the hemicylinder in the order of $10^{-4}$.

**Acknowledgement**

This work is supported by the State Committee for Science and Education of the Ministry of Education and Science of the Republic of Armenia (Grant No. 18T-1C114).

**References**

[1]     S. Rosenblum, Y. Lovsky, L. Arazi, F. Vollmer, B. Dayan, Cavity ring-up spectroscopy for ultrafast sensing with optical microresonators, Nat. Commun. 6 (2015) 6788. doi:10.1038/ncomms7788.

[2]     M.R. Foreman, J.D. Swaim, F. Vollmer, Whispering gallery mode sensors, Adv. Opt. Photonics. 7 (2015) 168. doi:10.1364/AOP.7.000168.

[3]     V.A. Nguyen, V.D. Pham, T.H.C. Hoang, H.T. Le, T.T. Hoang, Q.M. Ngo, V.H. Pham, A quantitative analysis of the whispering-gallery-mode lasers in Er3+-doped silica glass microspheres towards integr, Opt. Commun. 440 (2019) 14–20. doi:10.1016/j.optcom.2019.01.011.

[4]     W. Zhang, J. Yao, On-chip silicon photonic integrated frequency-tunable bandpass microwave photonic filter, Opt. Lett. 43 (2018) 3622. doi:10.1364/OL.43.003622.

[5]     A.E. Shitikov, I.A. Bilenko, N.M. Kondratiev, V.E. Lobanov, A. Markosyan, M.L. Gorodetsky, Billion Q-factor in silicon WGM resonators, Optica. 5 (2018) 1525. doi:10.1364/OPTICA.5.001525.

[6]     K.Y. Yang, D.Y. Oh, S.H. Lee, Q.-F. Yang, X. Yi, B. Shen, H. Wang, K. Vahala, Bridging ultrahigh-Q devices and photonic circuits, Nat. Photonics. 12 (2018) 297–302. doi:10.1038/s41566-018-0132-5.

[7]     M. Soltani, V. Ilchenko, A. Matsko, A. Savchenkov, J. Schlafer, C. Ryan, L. Maleki, Ultrahigh Q whispering gallery mode electro-optic resonators on a silicon photonic chip, Opt. Lett. 41 (2016) 4375. doi:10.1364/OL.41.004375.

[8] M. Förtsch, G. Schunk, J.U. Fürst, D. Strekalov, T. Gerrits, M.J. Stevens, F. Sedlmeir, H.G.L. Schwefel, S.W. Nam, G. Leuchs, C. Marquardt, Highly efficient generation of single-mode photon pairs from a crystalline whispering-gallery-mode resonator source, Phys. Rev. A. 91 (2015) 023812. doi:10.1103/PhysRevA.91.023812.

[9] Y. Wang, H. Li, L. Zhao, Y. Liu, S. Liu, J. Yang, Tapered optical fiber waveguide coupling to whispering gallery modes of liquid crystal microdroplet for thermal sensing application, Opt. Express. 25 (2017) 918. doi:10.1364/OE.25.000918.

[10] L. Cai, J. Pan, S. Hu, Overview of the coupling methods used in whispering gallery mode resonator systems for sensing, Opt. Lasers Eng. 127 (2020) 105968. doi:10.1016/j.optlaseng.2019.105968.

[11] A.B. Matsko, V.S. Ilchenko, Optical resonators with whispering-gallery modes-part I: basics, IEEE J. Sel. Top. Quantum Electron. 12 (2006) 3–14. doi:10.1109/JSTQE.2005.862952.

[12] C. Baker, C. Belacel, A. Andronico, P. Senellart, A. Lemaitre, E. Galopin, S. Ducci, G. Leo, I. Favero, Critical optical coupling between a GaAs disk and a nanowaveguide suspended on the chip, Appl. Phys. Lett. 99 (2011) 151117. doi:10.1063/1.3651493.

[13] M. Ghulinyan, R. Guider, G. Pucker, L. Pavesi, Monolithic Whispering-Gallery Mode Resonators With Vertically Coupled Integrated Bus Waveguides, IEEE Photonics Technol. Lett. 23 (2011) 1166–1168. doi:10.1109/LPT.2011.2157487.

[14] F. Ramiro-Manzano, N. Prtljaga, L. Pavesi, G. Pucker, M. Ghulinyan, A fully integrated high-Q Whispering-Gallery Wedge Resonator, Opt. Express. 20 (2012) 22934. doi:10.1364/OE.20.022934.

[15] H. Haroyan, H. Parsamyan, T. Yezekyan, K. Nerkararyan, Semicylindrical microresonator: excitation, modal structure, and Q-factor, Appl. Opt. 57 (2018) 6309. doi:10.1364/AO.57.006309.

[16] M. Cai, O. Painter, K.J. Vahala, Observation of Critical Coupling in a Fiber Taper to a Silica-Microsphere Whispering-Gallery Mode System, Phys. Rev. Lett. 85 (2000) 74–77. doi:10.1103/PhysRevLett.85.74.

[17] P.B. Johnson, R.W. Christy, Optical Constants of the Noble Metals, Phys. Rev. B. 6 (1972) 4370–4379. doi:10.1103/PhysRevB.6.4370.

[18] G.D. Pettit, W.J. Turner, Refractive Index of InP, J. Appl. Phys. 36 (1965) 2081–2081. doi:10.1063/1.1714410.

[19] I.H. Malitson, Interspecimen Comparison of the Refractive Index of Fused Silica*,†, J. Opt. Soc. Am. 55 (1965) 1205. doi:10.1364/JOSA.55.001205.

[20] S.A. Maier, Plasmonics: Fundamentals and Applications, Springer US, New York, NY, 2007. doi:10.1007/0-387-37825-1.

[21] R. Voelkel, Wafer-scale micro-optics fabrication, Adv. Opt. Technol. 1 (2012). doi:10.1515/aot-2012-0013.

[22] W. Yuan, L.-H. Li, W.-B. Lee, C.-Y. Chan, Fabrication of Microlens Array and Its Application: A Review, Chinese J. Mech. Eng. 31 (2018) 16. doi:10.1186/s10033-018-0204-y.

[23] E.M. Strzelecka, G.D. Robinson, L.A. Coldren, E.L. Hu, Fabrication of refractive microlenses in semiconductors by mask shape transfer in reactive ion etching, Microelectron. Eng. 35 (1997) 385–388. doi:10.1016/S0167-9317(96)00206-7.